\documentclass[conference]{IEEEtran}
\IEEEoverridecommandlockouts

\usepackage{cite}
\usepackage{amsmath,amssymb,amsfonts}
\usepackage{graphicx}
\usepackage{textcomp}
\usepackage{xcolor}
\usepackage{url}
\usepackage{hyperref}
\usepackage{booktabs}
\usepackage{tikz}
\usepackage{pgfplots}
\pgfplotsset{compat=newest}
\usepackage{microtype}
\usepackage{listings}
\graphicspath{{./}{figures/}}

\begin{document}
\title{A MATLAB toolbox for computation\\ of Speech Transmission Index (STI)
\thanks{Research described in this paper was supported by the Ministry of the Interior of the Czech Republic, program
IMPAKT1, under project VJ01010035 “Security risks of photonic communication networks”\!.}}

\author{\IEEEauthorblockN{Pavel Rajmic, Jiří Schimmel, and Šimon Cieslar}
\IEEEauthorblockA{\textit{%
Faculty of Electrical Engineering and Communication}\\
\textit{Brno University of Technology}\\
Brno, Czech Republic\\
pavel.rajmic@vut.cz}
}

\maketitle

\begin{abstract}
The speech transmission index (STI) is a~popular simple metric for the prediction of speech intelligibility when speech is passed through a transmission
channel.
Computation of STI from acoustic measurements is described in the \mbox{IEC\,60268-16:2020} standard.
Though, reliable implementations of STI are not publicly accessible and are frequently limited to the use with a~proprietary measurement hardware.
We present a~Matlab STI implementation of both the direct and indirect approaches according to the standard, including the shortened STIPA protocol.
The suggested implementation meets prescribed requirements, as evidenced by tests on reference signals.
Additionally, we conducted a~verification measurement in comparison to a~commercial measurement device.
Our software comes with open source code.
\end{abstract}

\begin{IEEEkeywords}
Speech transmission index, STI, STIPA, MATLAB toolbox, speech intelligibility.
\end{IEEEkeywords}


\section{Introduction}
Speech intelligibility refers to the extent of information conveyed in a transmitted speech signal, making it a crucial element in assessing the quality of speech signal transmitted through a transmission channel.
Various standards outline requirements for speech intelligibility that must be met and verified upon system installation, see for instance
\cite{ISO7240-19,NFPA72}.
This applies not only to purposes of emergency states in public address (PA) systems but also to environments such as lecture halls, theaters, etc., where speech intelligibility is important.


Standardized methodologies are available for the objective evaluation of speech codecs, including
ITU-T P.862 (PESQ) \cite{Rix2001:PESQ},
ITU-T P.863 (POLQA) \cite{ITU-T-P863:2018},
and ITU-T P.563 \cite{ITU-T-P563},
along with other techniques such as PEMO-Q \cite{Huber:2006a}
or ViSQOL \cite{Hines2015:ViSQOL}.
These are designed to gauge quality by concentrating on signal processing in voice coders and typical voice transmission issues within data networks.
Nonetheless, such problems are not directly related to the intelligibility of PA systems.
The room acoustics must be taken into account as well.
More recent approaches for evaluating intelligibility have been developed beyond STI, such as STOI (Short-Time Objective Intelligibility) \cite{Taal2011:Algor.Intelligib.Prediction,Jensen2016:Speech.Intelligib.Modulated.Mask},
but these have yet to be standardized.
The Speech Transmission Index (STI) serves as a well-established objective measure to evaluate the extent to which speech intelligibility is compromised after passing a~transmission channel. The STI for a specific channel is determined by comparing the signal measured at the output of the channel with the test input signal. The standardization of the STI quantification process is documented in \cite{sti_norm2020},
%
which is the fifth edition of the standard.

Although the STI/STIPA measurements are extensively used, achieving a~dependable, high-quality implementation
appears to be primarily within the domain of audio measurement equipment manufacturers.
STI/STIPA modules are either integrated or available for separate purchase in devices from companies like NTi Audio, Audio Precision, Brüel\&Kjær, Embedded Acoustics, Bedrock Audio.
This situation initially inspired the implementation of the STIPA method for public access. Currently, some open-source repositories are available online; however, none strictly adhere to the standard
\cite{sti_norm2020}. For instance, Jon Polom's GitHub repository%
\footnote{\url{https://github.com/jmpolom/sti-wav}}
offers an STI estimation model that utilizes real speech acquisition, but this does not conform to the standard.
The paper
\cite{Zaviska2024:STIPA} provides STI computation, but it limits itself only to the accelerated STIPA direct measurement.


Therefore, this paper offers an overview of the STI theory, subsequently applied in Matlab.
Our openly accessible code incorporates both direct and indirect methods as per the standard and enables STI estimation from both accelerated and original per-band acoustic data.
We fulfill the prescribed requirements, validated through comparison between our measurement and a~commercial device.

\section{Speech Transmission Index}
\label{sec:sti}

A speech signal passes through a transmission channel, such as a room, a telephone line, or an electro-acoustic channel consisting of a microphone, amplifier, and speaker. This transmission may include various types of signal processing, whether in analog or digital formats. The common applications and constraints of the STI model are detailed in \cite{sti_norm2020}.

The STI evaluates the physical aspects of the transmission channel such as noise and reverberation, and quantifies the ability of the channel to maintain speech intelligibility by assessing the variance between the input and output signals. The STI is represented as a real number from 0 to 1 used to quantify the degradation of speech intelligibility, with higher values indicating greater intelligibility. For instance, an STI of 0.58 is indicative of 'high-quality PA systems,' as found in venues such as concert halls and contemporary churches; at this STI level, complex messages should be easily understood by native speakers~\cite{sti_norm2020}.



The standard specifies two approaches for deriving the STI.
The \emph{direct method} involves the use of a speech-like test signal, while \emph{indirect method} depends on the system impulse response, and the standard does not prescribe any specific way to estimate it.
A basic summary is provided in Table~\ref{tab:DirectIndirectComparison};
for more properties, see Tables 3 and~4 in the standard~\cite{sti_norm2020}.
In the following,
we describe the fundamental core steps of each approach as outlined in the standard, along with comments on our actual implementation.
The Matlab source code is available at GitHub.\!\footnote{\url{https://github.com/rajmic/Speech-Transmission-Index-STI}}

\begin{table}
    \caption{Basic comparison of the direct and indirect methods~\cite{sti_norm2020}.}
    \label{tab:DirectIndirectComparison}%
    \vspace{-.5ex}%
    \centering
    \begin{tabular}{c|c|c}
        Property    & Direct method        & Indirect method\\
        \hline
        Channel type & General  & Linear, time-invariant\\
        Test signal & Modulated pink noise & Sweep/MLS\\
        Sensitivity to noise & Higher & Lower\\
        Accuracy    & Good & Excellent\\
        Demands     & Lower & Higher (precise device)\\
        Post-processing & Possible & Necessary
    \end{tabular}
\end{table}

The fundamental marker for deriving the STI is the observed reduction in modulation depth after the signal traverses the transmission path between the talker and the listener.
This reduction is assessed by comparing the input and output signals across various modulation frequencies ranging from 0.63 to 12.5\,Hz and multiple noise bands between 125 and 8\,000\,Hz. The data is subsequently consolidated into a~single STI value.
The just described process is common for all variants, both the direct and indirect class, but they differ by how the modulation depths are acquired and estimated.


\subsection{The direct method}
\label{sec:Direct.method}
\subsubsection{Input signal}
The synthetic input signal simulates the properties of speech signals. It accomplishes this by modulating broadband pink noise with amplitude modulations across several frequency bands. Long-term modulations represent words and sentences, while high-frequency modulations correspond to syllables at a rate of 3--4 per second. The benefits of using this type of signal include its simplicity, reproducibility, and language independence.

The Full STI represents a straightforward approach:
all combinations of 7 frequency bands and 14 modulation frequencies are used to form 98 test signals.
As long as the standard~\cite{sti_norm2020}
recommends a duration of at least 10 seconds per signal, the Full~STI procedure takes more than 15 minutes.
Such a duration can be prohibitive
(and is prone to disturbances). 
Therefore, a more concise approach has been designed to obtain STI, called STIPA.
The STIPA approach significantly accelerates the measurement at the cost of a acceptable error.
In a STIPA measurement, the input signal is formed such that only two selected modulation frequencies are used in each of the seven bands.
STIPA measurement takes 15--25~seconds.

To be more precise about the signal generation process,
a~noise generator and a set of seven filters
are the main components needed.
We used a 20th-order half-octave filter bank.

\begin{figure}[t] 
    \centering
    \input{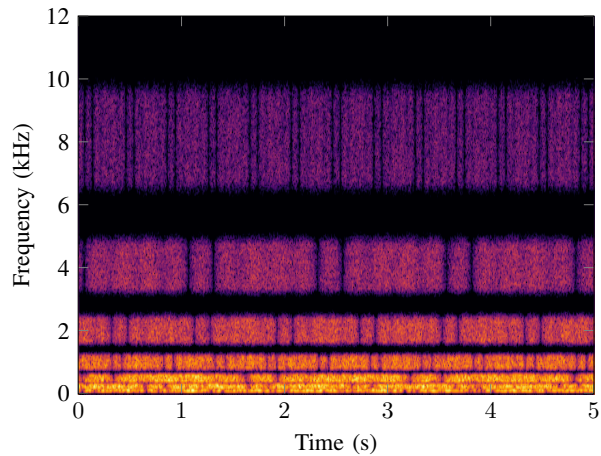}
    \vspace{-1em}
    \caption{Spectrogram of a 5-second-long excerpt of the STIPA test signal.}
    \label{fig:stipa_spectrogram}
\end{figure}

Formally, the set of Full STI signals is
\begin{equation}
\left\{ G_k N_k(t) A_m(t) \right\}_{k=1,\dots,7,\ m=1,\dots,14}
\end{equation}
where
$k$ is the octave band number, 
$G_k=10^{L_k/20}$ is the octave band weighting factor (levels $L_k$ in decibels are prescribed by the standard).
$N_k(t)$ is the band-limited noise-carrier signal and 
$A_m(t)$ is the amplitude modulator
\begin{equation}
    \label{eq:amp.modulation.FullSTI}
    A_m(t) = \sqrt{0.5\left(1+\cos(2\pi f_{m}t)\right)}.
\end{equation}
In contrast to this, the STIPA input signal is a single signal, a~mixture
\begin{equation}
\sum_{k=1}^7G_kN_k(t)A_k(t),
\end{equation}
with
\begin{equation}
\label{eq:STIPA.input.signal.modulations}
A_k(t) = \sqrt{0.5\left(1+0.55(\sin(2\pi f_{1k}t) - \sin(2\pi f_{2k}t))\right)}
\end{equation}
involving only two modulation frequencies
$f_{1k}, f_{2k}$ per band.
The factor of 0.55 is the modulation depth, common for all modulation frequencies and frequency bands.
Fig.\,\ref{fig:stipa_spectrogram} shows an excerpt of a~STIPA signal in the time-frequency plane.



\subsubsection{Measurement chain}
The respective test signals are inserted into the transmission channel, and the output is obtained in digital form.

\subsubsection{STI computation}
To obtain the STI, a number of steps must be followed on the recorded signal that are enumerated below.
The reductions of modulation depths have to be quantified, adjusted and mixed into a~single final number.

Comparing the two approaches, STIPA requires one more step before proceeding the rest of the computations, which is then shared with Full STI.
In particular, STIPA requires that the observed signal is first split into seven signals corresponding to the carrier noise bands.
The standard defines the filters rather generally but they must achieve a~minimum of 42~dB attenuation at the center frequency of each adjacent band~\cite{IEC_61260-1_octave_filters}.
We use filters of order 18, designed by the Matlab tool
\href{https://nl.mathworks.com/help/audio/ug/octave-band-and-fractional-octave-band-filters.html}{octaveFilter}.
The filter design
technique uses mapping the desired filter to a~Butterworth analog prototype, which is then mapped back to the digital domain
\cite{Orfanidis95:Intro.Signal.Proces}.
Based on our practical observations, we also cut off the first 200~ms of all resulting signals to safely avoid transient effects of the IIR octave filters.

The subsequent steps are identical for STIPA and Full STI.
\begin{enumerate}
    \item
        \textbf{Envelope detection}
        \label{itm:envelope.detection}
        -- the intensity envelope has to be determined by taking the square of the signal sample-wise, followed by a~low-pass filter with a cut-off frequency of approximately 100~Hz.
        For Full STI, the envelope is estimated from all 98 input signals,
        whereas in STIPA, this has to be done only on the 7 bandpass-filtered signals.
        We used the
        \href{https://nl.mathworks.com/help/signal/ref/lowpass.html}{lowpass} function of Matlab
        with precisely 100~Hz as the passband frequency.
        This procedure yields 98 envelopes $I_{k,m}(t)$ in the case of Full STI and 7~envelopes $I_k(t)$ in the case of STIPA.
    \item
        \label{itm:Calculate.modulation.depths}
        \textbf{Calculation of modulation depths} -- the modulation depths for each octave band and modulation frequency have to be estimated.
        Such procedure must be always carried out over a~whole number of periods for each modulation frequency,
        otherwise the estimation
        would be biased by the spectrum leakage.
        We achieve this by simply cutting off the suitable number of signal samples from the signal end.\\
        The modulation depth of the observed (output) signal is determined via computing
        \begin{gather*} 
            m_{\mathrm{o}}(k,f_m) = 
            \\
            \hspace*{-.45em}
            \frac{\raisebox{-.33ex}{2}
            \sqrt{\left[\sum_t\! I_{k,m}(t)\!\cdot\!\sin(2\pi f_m t)\right]^2\!\!+ \left[\sum_t\! I_{k,m}(t)\!\cdot\!\cos(2\pi f_m t)\right]^2}}{\sum_t I_{k,m}(t)}
        \end{gather*}
        for Full STI, resulting in 98~numbers.
        For STIPA, 14~estimates of the output depth are obtained using
        \begin{gather*} 
            m_{\mathrm{o}}(k,f_m) = 
            \\
            \hspace*{-.35em}
            \frac{\raisebox{-.33ex}{2}
            \sqrt{\left[\sum_t I_k(t)\!\cdot\!\sin(2\pi f_m t)\right]^2\!\!+ \left[\sum_t I_k(t)\!\cdot\!\cos(2\pi f_m t)\right]^2}}{\sum_t I_k(t)},
        \end{gather*}
        where $f_m$ should be substituted by $f_{1k}$ and $f_{2k}$, $k=1,\dots,7$,
        as in \eqref{eq:STIPA.input.signal.modulations}.
        The intensities $I_{k,m}(t)$ and $I_k(t)$ present in the calculations come from step \ref{itm:envelope.detection}.
    \item
        \textbf{Determination of modulation transfer ratios} --
        The modulation transfer ratio $m_{k,f_m}$ in band $k$ at the frequency modulation $f_m$ is calculated simply as the ratio of the output and input depths, i.e.,
        $$
        m_{k, f_m} = m_{\mathrm{o}}(k, f_m)/m_{\mathrm{i}}(k, f_m).
        $$
        %
        Our implementation allows fixing all of input $m_{\mathrm{i}}(k, f_m)$ to their nominal values.
        In the case of Full~STI, it is 1, and for STIPA, the nominal modulation depth is 0.55,
        see formulas \eqref{eq:amp.modulation.FullSTI} and \eqref{eq:STIPA.input.signal.modulations}, respectively.
        Such a~regime is default when no reference signal is passed to the 
        corresponding Matlab function.
        Otherwise, these input depths are calculated from the provided input signal, analogously to the above expression.

    \item
        \label{itm:Limit.mk}
        \textbf{Limiting modulation ratios} -- to avoid complex values in the SNR value  computed further, the modulation transfer values are limited to 1 if they exceed it:
        $$
        m_{k, f_m} = \mathrm{min}(m_{k, f_m}, 1).
        $$
    \item
        \label{itm:Take.ambient.noise.into.account}
        \textbf{Taking ambient noise into account} --
        when STI measurements are carried out in noiseless conditions, this step can predict the intelligibility index for the case of possible presence of ambient noise.
        In such a case, the characteristics of the ambient noise (or a multitude of them) is measured separately.
        The modulation transfer ratios from the preceding point are adjusted following
        $$
        m_{k, f_m} = m_{k, f_m} \cdot \frac{I_{\textup{s},k}}{I_{\textup{s},k}+I_{\textup{n},k}}.
        $$
        Here, $I_{\textup{s},k}$ represents the acoustic intensity of the test signal in band $k$,
        and $I_{\textup{n},k}$ is the intensity of ambient noise in the corresponding band.\\
        In our implementation, this step is performed when both vectors of test signal levels and ambient noise levels in octave bands are provided.
        Otherwise, no adjustment is done 
        and coefficients $m_{k, f_m}$ are just retrieved from step~\ref{itm:Limit.mk}.
    \item
        \label{itm:Taking.auditory.effects.into.account}
        \textbf{Taking auditory effects into account} --
        since frequency-dependent auditory effects take place in real situations, the STI methodology
        incorporates these effects in the form of a~reduction of the modulation transfer ratios. 
        Auditory effects are taken into account only when the signal is obtained acoustically and when the total intensity in octave bands are known.
        They are adjusted according to
        $$
        m_{k, f_m} = m_{k, f_m} \cdot \frac{I_k}{I_k+I_{\textup{am},k}+I_{\textup{rt},k}}.
        $$
        Here, $I_k=I_{\textup{s},k}+I_{\textup{n},k}$ denotes the total acoustic intensity discussed in step \ref{itm:Take.ambient.noise.into.account}.
        The term $I_{\textup{am},k}$ stands for the auditory masking in octave band $k$ and is computed via
        $$
            I_{\textup{am},k} = 10^{L_{k-1}/10} \cdot 10^{L_{\textup{a},k}/10}
        $$
        where $L_{k-1}$ is the intensity in band $k-1$,
        and $L_{\textup{a},k}$ are given by Table A.2 in the standard \cite{sti_norm2020}.
        Masking effects are not modeled in
        band $k=1$.

        Further on, $I_{\textup{rt},k}$ takes into account the absolute reception threshold, which involves the absolute threshold of hearing and the minimal required dynamic range for a~correct recognition of speech.
        This quantity is again frequency-dependent,
        and
        $$
            I_{\textup{rt},k}= 10^{A_k/10}
        $$
        is determined by coefficients $A_k$ from Table A.3 of the standard.\\
        In our implementation, step
        \ref{itm:Taking.auditory.effects.into.account} is performed when the vector of the octave-band signal levels is provided.
    \item
        \textbf{SNR computation} -- the value of the effective SNR is computed from the limited modulation transfer ratios,
        $$\mathit{SNR}^{\mathrm{eff}}_{k, f_m} = 10 \log_{10}{\frac{m_{k, f_m}}{1-m_{k, f_m}}},
        $$
        and the result is limited such as not to exceed the range of $[-15, 15]$~dB.
    \item
        \label{itm:TI.calculation}
        \textbf{Transmission indexes determination} -- this index is determined for the SNR value in each band:
        $$\mathit{TI}_{k, f_m} = \frac{\mathit{SNR}^{\mathrm{eff}}_{k, f_m}+15}{30}.$$
        Clearly, such an index resides in the interval $[0, 1]$.
    \item
        \textbf{Modulation Transfer index (MTI)} -- the MTI of each band is computed via taking an unweighted average value over the frequencies:
        $$M_k = \mathit{MTI}_k = \frac{1}{n}\sum_{m=1}^n \mathit{TI}_{k, f_m}.$$
        In the expression, we have $n=14$ for Full STI, while in the case of STIPA $n=2$.
        After this step, Full STI and STIPA both remain with seven MTI values.
    \item
        \textbf{STI computation} -- calculate the final value of the Speech transmission index as 
        $$\mathit{STI} = \sum_{k=1}^7 \alpha_k M_k - \sum_{k=1}^6 \beta_k \sqrt{M_k \cdot M_{k+1}},$$
        where the first part STI takes into account the intra-band modulations and the second part depends on MTIs of adjacent bands.
        The factors $\alpha_k$, $\beta_k$ in the expression are gender-specific factors for octave band $k$,
        given in Table~A.1 of the standard
        \cite{sti_norm2020}.
        Only male-related coefficients should be used, which corresponds to this generally worse intelligibility scenario.
        In the event that STI is greater than one, the result is clipped to that value.
\end{enumerate}

\subsection{The indirect method}
In this method, the indicators of reduction of modulation ratios
(and thus of decreased intelligibility) are derived indirectly,
through an estimate of the impulse response of the system.
Clearly, this assumes that the transmission system is linear and time-invariant.

\subsubsection{Input signal}
One of the most popular methods how to acquire the system impulse response is using a swept-sine signal at the input of the system.
In particular, for the impulse response determination, we use the method described in
\cite{Farina2000:Impulse.response.measurement.swept.sine}
in our implementation,
but any other approach or tool can be used.
Note that the standard
\cite{sti_norm2020}
requires the sweeps no shorter than 1.6 seconds.

\subsubsection{Measurement chain}
Once the sweep signal has passed through the system and has been acquired, 
the method of 
\cite{Farina2000:Impulse.response.measurement.swept.sine}
suggests estimation of the impulse response by a~deconvolution with an inverse-sweep filter.

\subsubsection{STI computation}
The obtained impulse response is now filtered such as to obtain seven frequency bands as with the direct method.
It is done with exactly the same filters as described in Section \ref{sec:Direct.method}.
Now, the seven filtered impulse responses, $h_k(t)$, 
are plugged into the following expression, often called the Schroeder equation~\cite{Schroeder1981:MTF},
for the determination of modulation transfer ratios:
\begin{equation}
    \label{eq:Schroeder.for.indirect}
    m_{k, f_m} \!=
    \frac{%
    \left|\sum_{t} h_k^2(t) \exp(-\textup{j}2\pi f_m t) \right|}%
    {%
    \sum_{t} h_k^2(t)
    }
    \cdot
    [1+10^{-\mathit{SNR_{k}}/10}]^{-1}
    \end{equation}
where $\mathit{SNR_{k}}$ is the signal to noise ratio in band $k$.
For Full~STI, we have 98 such ratios, while for STIPA there is 14 of them
(note, however, that computing the compact STIPA yet having access to all 98 ratios can rarely make sense).

From this point on, the process is identical to the direct method, i.e., step~\ref{itm:Limit.mk} and so on
are followed until the STI is obtained.
In our implementation, taking $\mathit{SNR_{k}}$ into account is not performed right in \eqref{eq:Schroeder.for.indirect}
but rather in step \ref{itm:Take.ambient.noise.into.account} of the computation;
for the equivalence of the two options, note that it holds $\mathit{SNR_{k}}=10\log(\mathit{I_{s,k}}/\mathit{I_{n,k}})$.

\section{Implementation}

The code has been developed in Matlab version 2023a. It uses a~few functions of the Signal processing toolbox and of the Audio toolbox introduced in version 2018a at the latest.

The toolbox was implemented in the form of separate functions for calculating the STI:
one using the direct Full STI, one using the direct STIPA methods and the other one using the indirect method from an impulse response:
\begin{lstlisting}
STI = stipa(sig,fs)
STI = fullsti(sig,fs)
STI = sti_ir(sig,fs)    
\end{lstlisting}
Here, \texttt{sig} is the received signal and \texttt{fs} is the sampling frequency.
To access to the optional parts of the STI calculation such as the correction for ambient noise, the user should use the named parameters and the input parser scheme introduced in Matlab 2007a.
Details about the input and output parameters of the above functions can be found on our GitHub.

The toolbox includes functions for generating test signals for the Full STI and STIPA methods, and a~swept-sine signal as proposed in 
\cite{Farina2000:Impulse.response.measurement.swept.sine}:
\begin{lstlisting}
sig = generateStipaSignal(duration,fs)
sig = generateFullSTISignal(duration,fs)
sig = generateSweptSineSignal(duration,fs)
\end{lstlisting}
Here, \texttt{sig} is the generated signal and \texttt{duration} is its duration in seconds.
In the case of the Full STI signal, it represents the duration of each of the 98 modulated signals.
Both generated Full STI and STIPA signals are normalized to an RMS value such that the signal does not clip when saving to a~\texttt{wav} file.
The generating functions also have optional parameters; see the GitHub documentation for details. 

The toolbox also includes scripts for demonstration of the above described functions. The names of these scripts start with the \texttt{demonstration} prefix.

\begin{figure}[t]
    \centering
    \includegraphics[width=.99\linewidth]{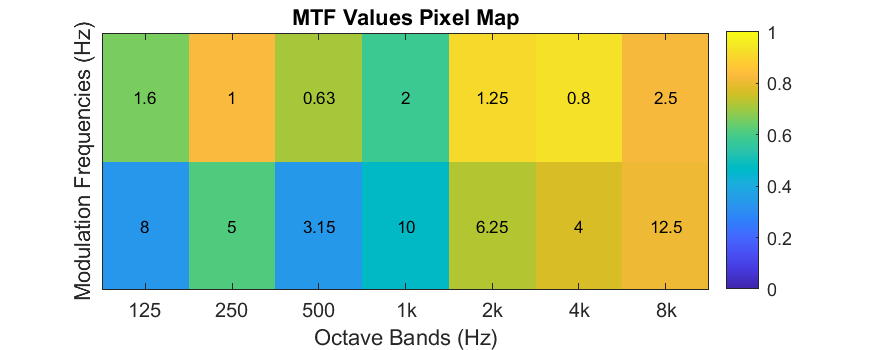}%
    \vspace*{-1.7ex}%
    \caption{Example of the pixel map of the 14 STIPA modulation indexes}
    \label{fig:Example.MTF}
\end{figure}

One of the goals of the implementation was to demonstrate the STI calculation algorithm so that it could be used in teaching.
Therefore, the implementation is divided into multiple functions following the individual calculation steps listed in Section \ref{sec:sti}, and also includes the ability to visualize the output of several intermediate steps of the  algorithm:

\begin{itemize}
\item Text panel with the resulting STI value and the corresponding qualification category.
\item Pixel map of modulation transfer ratios:
shows a 2D matrix of these ratios for different modulation frequencies and octave bands, see Fig.~\ref{fig:Example.MTF},
as computed in step
\ref{itm:TI.calculation}.
\item MTI bar graph: shows MTI values for each band. 
\item SPL graphs: these show the sound pressure levels and A-weighted sound pressure levels in the bands. If auditory masking analysis is active, additional curves are added: “Masking” (noise level), “Threshold” (threshold level) and “Total S+N” (total signal+noise level).
\item A table with MTF values, MTI and weighting/redundancy factors, STI, and others. 
\end{itemize}

The toolbox also includes
\texttt {verificationTests.m}, a~script that verifies the implementation using test signals described in Annex~C of the standard \cite{sti_norm2020}.
The results are described in \cite{Zaviska2024:STIPA}.
The script that requires test signals developed by Embedded Acoustics%
\footnote{\url{http://www.stipa.info/index.php/download-test-signals}}.

\section{Verification Measurement} 
\label{sec:measurements}

The Matlab implementation was verified in a real environment.
The direct Full~STI, STIPA, and the indirect Full~STI results were compared against each other, and against a~professional measurement device as a~reference.

\subsection{Venue}
The verification measurement took place in the lecture room
at the Faculty of Electrical Engineering and Communication, Brno University of Technology.
It is a room with a capacity of ca 200 students, ca 17~meters long and 12~meters wide.
There are no windows, reducing noise coming from outside.
There were only two persons in the room.
The background noise was low, with the maximum sound level of 20\,dB(A) in the octave band with the center frequency of 4 kHz.
Yet, upon taking the measurement, a number of impulsive disturbances emerged (steps and closing the door in the adjacent corridor).

\begin{figure*}
    \centering
    \includegraphics[width=.7\linewidth]{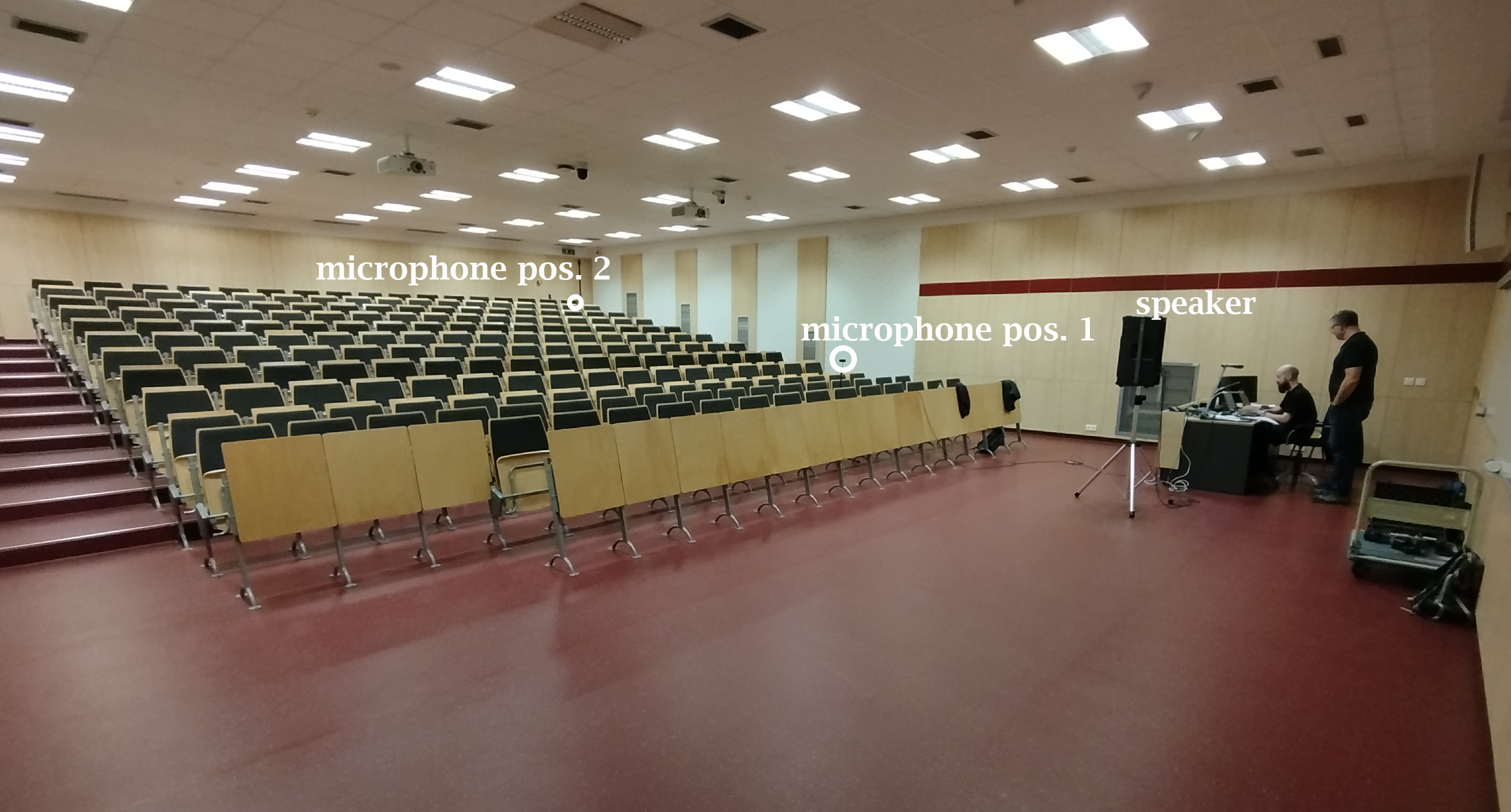}
    \caption{Loudspeaker and microphone placement in the auditorium.}
    \label{fig:measurement_setup}
\end{figure*}

\begin{figure}[t]
    \centering
    \vspace*{-5mm}
    \includegraphics[width=.9\linewidth]{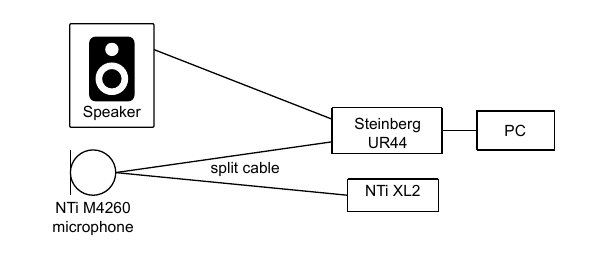}
    \vspace*{-5mm}
    \caption{Measurement setup used in verification}
    \label{fig:Measurement.schematic}
\end{figure}

\subsection{Setup}
The following devices were used in the measurement process:
\begin{itemize}
    \item B\&K Type 4231 calibrator,
    \item NTi Audio M4260
    measurement microphone,
    \item Yamaha MSR400 active speaker (no
    corrections),
    \item Audio interface Steinberg UR44
    connected to a laptop,
    \item Software on the laptop:
    Cubase 10 set to 24 bit/48 kHz, MATLAB R2024b, Room EQ Wizard (REW) ver.\,5.31.3,
    EASERA ver.\,1.2.18.17,
    \item NTi Audio XL2 as the reference
    STIPA analyzer.
\end{itemize}

The loudspeaker was positioned at a typical location of a~standing teacher,
and its volume was adjusted as to achieve a~“normal” level of 60~dB(A) at a~distance of 1~meter from the loudspeaker,
in line with the recommendation in
\cite{ISO_9921:2003,sti_norm2020}.

The input test signals were generated in Matlab with our tool, sampled at 48~kHz with 24~bit quantization depth.
Test signals were played back by Cubase (which acquired the microphone signal at the same time, see below).
The REW software was used as an alternative means of obtaining the impulse response.
EASERA was another off-the-shelf tool used to determine STI from the measured impulse responses.


After calibration, the same set of measurements was taken at two distinct microphone positions 
(3.4 and 12.5 meters away from the loudspeaker, corresponding to a person seated in the second and twelveth row of chairs, respectively).
 
The signal captured by the NTi Audio M4260 microphone was was routed to NTi Audio XL2 audio analyzer, which measured the sound pressure level in dB(A).
The internal implementation of STIPA in the NTi analyzer has also been used a~reference.
Simultaneously, the signal from the microphone was recorded using a~laptop with Cubase and the UR44 audio interface.
Before the acquired signals were imported to Matlab,
they were checked for clipping and the border parts with no useful signal were cut out.

A~schematic of the verification setup is in Fig.~\ref{fig:Measurement.schematic};
an example setup of the loudspeaker and the microphone placement can be seen in Fig.~\ref{fig:measurement_setup}.
Characteristics of the UR44 were verified using the Audio Precision APx525 analyzer; for details, see~\cite{Zaviska2024:STIPA}.

\subsection{Results}
Table~\ref{tab:sti.values.position1} shows the obtained STIs  at microphone position~1 (i.e., closer to the speaker).
The Full STI was performed only once due to its duration of more than 15~minutes.
In contrast to that, STIPA measurement of 25~seconds in length was repeated.
Our indirect method was also carried out three times.

\begin{table}[b]
    \caption{STI values at position 1 (rounded to the second decimal)}
    \label{tab:sti.values.position1}%
    \vspace{-.5ex}%
    \centering
    \begin{tabular}{l||c||c|c|c}
        \hline
        \textbf{Measurement type} & \textbf{Mean} & \textbf{1} & \textbf{2} & \textbf{3} \\
        \hline\hline
        Full STI ours & 0.74 & 0.74 & --- & --- \\ \hline\hline
        STIPA ours & 0.72 & 0.72 & 0.71 & 0.72 \\
        \hline
        STIPA NTi XL2 & 0.70 & 0.70 & 0.70 & 0.70  \\ \hline\hline
        Indirect ours (full) & 0.75 & 0.75 & 0.76 & 0.75 \\ \hline
        Indirect REW (full) & 0.77 & 0.77 & --- & --- \\ \hline
        Indirect EASERA (full) & 0.79 & 0.79 & --- & ---\\ \hline
    \end{tabular}
\end{table}

The reference value from NTi XL2 STIPA was 0.70.
Our STI values were generally slightly higher, as seen in the table.
Given the fact that STIPA uncertainity is usually considered 0.03, and that STIPA itself can be less precise than the Full~STI, all the values can be accepted as reliable.
Note that both indirect approaches tend to produce the highest STI values.

Table~\ref{tab:sti.values.position2} presents the STI indexes from the microphone moved to the more distant position~2.
The reference value from NTi XL2 STIPA was 0.63.
Our STIPA value was similar, with the mean of 0.65.
But the Full STI and the indirect methods provide values that are significantly higher.
Therefore, for the first sight, STIPA approach(es) seem to be more sensitive to worsening of the listening conditions, compared to the rest of our procedures.
But we explain greater STI differences in position 2 by the impulsive disturbances encountered during the direct Full STI measurement.
And, the strikingly low values of STIPA indexes (notice that the two of them are in alignment)
are most likely to be explained by the standard~\cite{sti_norm2020} itself:
its Table~3 specifies that STIPA measurements are not suitable for situations involving echo.
Note that in contrast, the echo was masked by the direct acoustic wave in microphone position 1.

\begin{table}[t]
    \caption{STI values at position 2 (rounded to the second decimal)}
    \label{tab:sti.values.position2}%
    \vspace{-.5ex}%
    \centering
    \begin{tabular}{l||c||c|c|c|c|c}
        \hline
        \textbf{Measurement type} & \textbf{Mean} & \textbf{1} & \textbf{2} & \textbf{3} \\
        \hline\hline
        Full STI ours& 0.72 & 0.72 & --- & --- \\ \hline\hline
        STIPA ours & 0.65 & 0.66 & 0.65 & 0.65 \\ \hline
        STIPA NTi XL2 & 0.63 & 0.63 & 0.63 & 0.63 \\ \hline\hline
         Indirect ours (full) & 0.73 & 0.73 & 0.73 & 0.73 \\ \hline
         Indirect REW (full) & 0.77 & 0.77 & --- & --- \\ \hline
        Indirect EASERA (full) & 0.76 & 0.76 & --- & ---\\ \hline
    \end{tabular}
\end{table}

\section{Conclusion}
\label{sec:conclusion}

An open-source Matlab implementation of different STI estimators
was presented.
It closely follows the
standard
\cite{sti_norm2020}
and it has been numerically verified.
Our implementation, independent of a~hardware measurement device, can widen the range of possible applications.
Our code is also potentially valuable in education
thanks to the visualization features.
Other authors are invited to contribute to the code, and possibly test our implementation in various measurement scenarios.



\newcommand{\noopsort}[1]{} \newcommand{\printfirst}[2]{#1} \newcommand{\singleletter}[1]{#1} \newcommand{\switchargs}[2]{#2#1}

\end{document}